\documentstyle[11pt,aas2pp4]{article}

\def\masl{\ifmmode  {\rm M_{\sun}yr^{-1}} \else ${\rm M_{\sun}yr^{-1}}$\fi}
\def\massloss#1#2{$ #1\, 10^{#2}\,{\rm M_{\sun}yr^{-1}} $}
\def\mdot{\ifmmode  \dot{M} \else $\dot{M}$\fi}
\def\msun{\ifmmode M_{\odot} \else $M_{\odot}$\fi}
\def\vinf{\ifmmode v_{\infty} \else $v_{\infty}$\fi}
\def\teff{\ifmmode T_{\rm eff} \else $T_{\rm eff}$\fi}
\def\logg{\ifmmode \log g \else $\log g$\fi}
\def\loggeff{\ifmmode \log g_{\rm eff} \else $\log g_{\rm eff}$\fi}
\def\rstar{\ifmmode R_{\star} \else $R_{\star}$\fi}
\def\tstar{\ifmmode T_{\star} \else $T_{\star}$\fi}
\def\lstar{\ifmmode L_{\star} \else $L_{\star}$\fi}
\def\mstar{\ifmmode M_{\star} \else $M_{\star}$\fi}
\def\rsun{\ifmmode R_{\odot} \else $R_{\odot}$\fi}
\def\lsun{\ifmmode L_{\odot} \else $L_{\odot}$\fi}
\def\12c16o{$^{12}{\rm C}\left(\alpha,\gamma\right)^{16}{\rm O}$}
\def\kms{\ifmmode {\rm km \;s^{-1}} \else $\rm km \;s^{-1}$\fi}
\def\hei{He~{\sc i}}
\def\heii{He~{\sc ii}}
\def\gv{$\gamma^2$ Vel }
\def\zp{$\zeta$ Pup }

\def\apj{ApJ }
\def\apjs{ApJS }

\def\aj{AJ }
\def\aa{A\&A }
\def\aas{A\&AS }

\def\mnras{MNRAS }

\begin{document}

\title{Fundamental stellar parameters of $\zeta$ Pup and $\gamma^2$ Vel
	from HIPPARCOS data}

\author{Daniel Schaerer\altaffilmark{1,2,3}}
\altaffiltext{1}{Space Telescope Science Institute, 3700 San Martin Drive, 
Baltimore, MD 21218; schaerer@stsci.edu}
\altaffiltext{2}{Observatoire de Gen\`eve, CH-1290 Sauverny, Switzerland}
\altaffiltext{3}{Observatoire Midi-Pyr\'en\'ees, 14, Av.\ E.\ Belin, F-31400 
Toulouse, France}

\author{Werner Schmutz\altaffilmark{4}}
\altaffiltext{4}{Institut f\"ur Astronomie, ETH-Zentrum. CH-8092 Z\"urich, 
Switzerland}
\author{Michel Grenon\altaffilmark{2}}

\begin{abstract} 
We report parallax measurements by the HIPPARCOS satellite 
of $\zeta$ Puppis and $\gamma^2$ Velorum.
The distance of \zp is $d=429^{+120}_{-77}$~pc, in agreement
with the commonly adopted value to Vela OB2. 
However, a significantly smaller distance is found for the \gv system:
$d=258^{+41}_{-31}$~pc.

The total mass of \gv derived from its parallax, the angular size
of the semi-major axis as measured with intensity interferometry, 
and the period is $M(WR+O)=29.5 \pm 15.9 \msun$.
This result favors the orbital solution of Pike et al.\ \markcite{p83}
(1983) over that of Moffat et al.\ \markcite{m86} (1986).
The stellar parameters for the O star companion derived from 
line blanketed non-LTE atmosphere models are: $\teff=34000 \pm 1500$ K, 
$\log L/\lsun=5.3 \pm 0.15$ from 
which an evolutionary mass of $M=29 \pm 4 \, \msun$ and an age of 
$4.0^{+0.8}_{-0.5}$ Myr is obtained from single star evolutionary models.
With non-LTE model calculations including He and C we derive a luminosity
$\log L/\lsun \sim 4.7 \pm 0.2$ for the WR star.
The mass-luminosity relation of hydrogen-free WR stars 
implies a mass of $M_{\rm WR} \sim 5 \pm 1.5 \, \msun$.

From our data we favor an age of $\sim$ 10 Myr for the bulk of the 
Vela OB2 stars. 
Evolutionary scenarios for \zp and \gv are discussed
in the light of our results.


\centerline{\sl Submitted to ApJ Letter}

\end{abstract}

\twocolumn

\section{Introduction}
Accurate stellar parameters of massive stars are of fundamental
importance to guide our understanding of stellar evolution.
The closest single O star is \zp (HD 66811, O4I(n)f), which is used as a
prototype for radiation-driven wind models (e.g.\ Pauldrach et al.\ 
\markcite{p94} 1994, Schaerer \& Schmutz \markcite{ss94} 1994 for recent 
models).
The closest O and WR stars are found in the system $\gamma^2$ Vel
(HD\, 68273).
Its binary nature hampers a spectroscopic analysis of its
components but it has the advantage to allow direct mass determinations.

With the HIPPARCOS satellite it has been possible to obtain 
annual parallaxes for these two objects. In the present publication 
we discuss their fundamental stellar parameters based on the
HIPPARCOS distances. 
Establishing a reliable distance -- and hence 
luminosity -- for \zp is of particular
interest for quantitative comparisons of hydrodynamic wind models with
observations and the understanding of its observed He enrichment
(cf.\ Bohannan et al.\ \markcite{b90} 1990).
With an accurate distance to the \gv system we are in a position to 
test the mass-luminosity relation for the WR star.

Stellar parameters for \zp and \gv based on the HIPPARCOS parallaxes 
are revised and/or determined from non-LTE atmosphere models in \S 2.
In \S 3 we discuss evolutionary scenarios for these objects in 
the light of previous observations and new HIPPARCOS parallaxes
of few presumed members of the Vela OB2 and Vela R2 associations.

\section{Revised stellar parameters}

\subsection{\zp}
The HIPPARCOS parallax of \zp is 2.33 $\pm$ 0.51 mas corresponding
to a distance of $d=429^{+120}_{-77}$ pc.
This value is consistent with the canonically adopted value
of $d=$ 450 pc based on its presumed association with Vela
OB2 (cf.\ Brandt et al.\ \markcite{b71} 1971), which may, however, 
be questionable (see \S \ref{s_discussion}).
Distance dependent stellar parameters derived so far for \zp are 
thus compatible with the HIPPARCOS measurement.
We therefore only provide an update of these values adopting
the new distance measurement and its error.

Rescaled to the new distance the radio mass-loss rate from Lamers 
\& Leitherer \markcite{ll93} (1993) is $\log\mdot=-5.57 \pm 0.15$, where 
we adopt the same uncertainty as these authors.
As mentioned earlier the basic parameters of O stars with important 
winds are still somewhat uncertain (for \zp cf.\ Bohannan et al.\ 
\markcite{b90} 1990, Gabler et al.\ \markcite{g89} 1989, Schaerer \& 
Schmutz \markcite{ss94} 1994, Puls et al.\ \markcite{p96} 1996).
Adopting the model results (\teff=42000 K, BC) and observational parameter 
(V, A$_{\rm V}$) from Bohannan et al.\ \markcite{b90} (1990) the new 
distance yields the following values (the uncertainties reflect only 
the distance):
$\log L/\lsun= 5.87^{+0.25}_{-0.13}$ and $R/\rsun= 16.3^{+5.4}_{-2.3}$.
This value is compatible with the corrected angular diameter 
($\alpha = 4\arcsec .0 \, 10^{-4}$) from Kudritzki et al.\ \markcite{k83}
(1983) which gives $R/\rsun= 18.4^{+5.2}_{-3.3}$.
The Lyman continuum flux of \zp derived from recent models 
is $\log Q_0=49.7$ photon $s^{-1}$ (Bohannan et al.\ \markcite{b90} 1990;
Schaerer \& Schmutz \markcite{ss94} 1994).

\subsection{\gv}
Interestingly, \gv shows a significantly larger parallax than expected:
$\pi= 3.88 \pm 0.53$ mas, corresponding to $d=258^{+41}_{-31}$ pc.
The HIPPARCOS parallax could be influenced by the orbital motion.
By adopting the orbital orientation from Moffat et al.\ \markcite{m86} 
(1986), the inclination from St.-Louis et al.\ \markcite{stn87} (1987), 
the WR/O star luminosity ratio from Conti \& Smith \markcite{cs72} (1972), 
and the angular size of the semi-major axis from Hanbury Brown et al.\ 
\markcite{hb70} (1970) we find that the center of light is displaced by 
0.3 to 1.2 mas from its average location. 
An interpretation of the orbital motion as a standard 
deviation yields 0.8 mas. 
For smaller magnitude differences (cf.\ below) or the orbital parameters
of Pike et al.\ \markcite{p83} (1983) this error becomes smaller. 
The contribution of the orbital motion
to the error of the  HIPPARCOS parallax is $0.8/\sqrt{N}=0.09$ mas, 
where N=78
is the number of measurements used minus the number of fitted parameters.
The measured parallax of the spectroscopic binary system \gv provides 
therefore an accurate distance indicator. In the following we shall 
adopt the internal HIPPARCOS error as the uncertainty on the parallax.

Combining the HIPPARCOS parallax with the observed angular size of the 
semi-major axis from the interferometric measurements of Hanbury Brown 
et al.\ \markcite{hb70} (1970) and the known orbital period (cf. Niemela 
\& Sahade \markcite{ns80} 1980, Pike et al.\ \markcite{p83} 1983)
allows to derive the total mass of the system: 
$M(WR+O)=29.5 \pm 15.9 \msun$ (see Table \ref{ta_summary} for parameters).

In the recent literature, there are two different amplitudes of the orbital 
motion. Pike et al.\ \markcite{p83} (1983) derived a mass ratio of 
$q=0.36 \pm 0.09$, whereas Moffat et al.\ \markcite{m86} (1986) find a
significantly larger mass ratio of 
$q=0.54 \pm 0.03$, due to a larger radial velocity amplitude for the O star. 
Adopting an inclination of $i=70 ^\circ$ later confirmed by 
St.-Louis et al.\ \markcite{stn87} (1987), Pike et al.\ \markcite{p83} obtain 
$M(WR)=7.7 \pm 2.5 \msun$ and $M(O)=22 \pm 5 \msun$, 
while the Moffat et al.\ \markcite{m86} data yields $M(WR)=19^{+7}_{-2} \msun$
and $M(O)=35^{+13}_{-3} \msun$ (see Smith \& Maeder \markcite{sm89} 1989).
The total mass derived above using the HIPPARCOS parallax favors the
solution of Pike et al.\ \markcite{p83} (1983). 
On the other hand for the orbital parameters of Moffat et al.\ \markcite{m86}
(1986) the observed angular size of the semi-major axis would imply a 
distance of $d=308 \pm 37$ pc, which is within 1 $\sigma$ of the HIPPARCOS
distance. We conclude that the orbital parameters from both Pike et al.\
\markcite{p83} (1983) and Moffat et al.\ \markcite{m86} (1986) are compatible 
with the HIPPARCOS parallax and the
interferometric determination of the semi-major axis from Hanbury Brown et 
al.\ \markcite{hb70} (1970). The discrepancy between the two orbital 
solutions, however, implies a significant uncertainty for the masses of the 
\gv components.

{\footnotesize
\begin{table}[htb]
\caption{Derived absolute magnitudes for the \gv system for different
assumptions on the WR-O magnitude difference (CS: Conti \& Smith 
1972, BC93: Brownsberger \& Conti 1993).
$M_{\rm v}({\rm system})=-5.49$, $M_{\rm V}({\rm system})=-5.39$,
where $v$ stands for the Smith and $V$ the Johnson system}
\centerline{
\begin{tabular}{lrrrrrrrrrr}
\\ \hline \\
ID  & $\Delta M_{\rm V}$ & $M_{\rm v}$(WR) & $M_{\rm V}$(O) \\
\\ \hline
CS72 & 1.4  & -3.83 & -5.13 \\
BC93 & 1.9  & -3.42 & -5.22 \\
1	& 1.36	& -3.86	& -5.12 \\
2	& 1.75	& -3.54	& -5.20	\\
\hline
\end{tabular}
}
\label{ta_mv}
\end{table}
} 

The HIPPARCOS distance measurement places \gv considerably closer than
the previously adopted distance ($d \sim 450$ pc) based on its 
assumed association with Vela OB2 (e.g.\ Brandt et al.\ \markcite{b71} 1971, 
Sahu \markcite{s92} 1992). 
The apparent magnitude and extinction from the catalogue of van der Hucht
et al.\ \markcite{vdh88} (1988), the flux zero-point correction of Schmutz 
\& Vacca \markcite{sv91} (1991) and the revised distance yields
$M_v$(system)$=-5.49$.
Absolute magnitudes obtained for the WR and O star components,
assuming different luminosity ratios, are given in Table 
\ref{ta_mv}. The values of $\Delta M_V=1.4$~mag and 1.9~mag result from 
the comparison of equivalent widths of \gv with those from the WC8 star 
WR135 (Conti \& Smith \markcite{cs72} 1972; Brownsberger \& Conti 
\markcite{bc93} 1993).
The absolute brightness of the WC8 star is considerably 
smaller than the average value for WC8 stars given by van der 
Hucht et al.\ \markcite{vdh88} (1988) ($<M_v({\rm WC8})>=-4.8$). 

The bolometric correction of the WC8 star can be obtained from
evolutionary or atmosphere models. Adopting the mass-luminosity
relation of Schaerer \& Maeder \markcite{sm92} (1992) or Smith et al.\ 
\markcite{smm94} (1994) one  has $\log L/\lsun=5.06$ (5.69) for the 
low (high) mass and hence a  bolometric correction of BC=-4.2 to -3.7 
(-5.8 to -5.3).
Smith et al.\ \markcite{smm94} derived a value of -4.5 for this subtype.
From non-LTE atmosphere models (cf.\ below) we obtain BC=-3.5 to -3.4.

The brightness of the O star is typical for a giant 
(cf.\ Vacca et al.\ \markcite{v96} 1996).
Conti \& Smith \markcite{cs72} (1972) concluded that the O star is a 
supergiant, based on the ratio 4089 Si\,{\sc iv}/4143 He\,{\sc i}. 
However, they
note that the measurements have been difficult due to blending effects.
The fast rotation of the O star makes it indeed difficult to locate a weak
He\,{\sc i} line in a spectrum dominated by WR features and to
isolate Si\,{\sc iv} $\lambda 4089$ from a nearby nitrogen line.
Therefore, the spectroscopic luminosity classification is not severely
in conflict with the absolute magnitude of the O star.
Comparing two spectra from Kaufer et al.\ \markcite{k97} (1997) observed 
at the phases $\phi=0.24$ and 0.72 we find that He{\sc ii} $\lambda 4686$
 is in absorption. Its equivalent width corrected for the emission of
the WR star is compatible with a classification as a giant (Mathys 
\markcite{m88} 1988), especially when considering the large uncertainty 
of measuring a broad absorption superimposed on a strong emission line. 
A statistical argument against a luminosity class I can be 
made based on the observed rotation of $v_{\rm rot} \sin\,i = 220$ \kms\ 
(Baade et al.\ \markcite{bsk90} 1990). For a supergiant such a velocity is
extremely rare whereas for giants such high values are more ordinary
(cf.\ Howarth et al.\ \markcite{h97} 1997).
We determine a 4471 He~{\sc i}/4541 He~{\sc ii} ratio of $1.6\pm 0.2$.
According to Mathys \markcite{m88} (1988) this classifies the O star as 
O8 or O8.5. 
Conti \& Smith \markcite{cs72} (1972) have determined O9 spectral type. 
We agree in the equivalent width for the 4471 line, but in our spectra,
the He{\sc ii} $\lambda 4541$ absorption is much stronger than what 
is cited by Conti \& Smith \markcite{cs72} (1972). 
We therefore favor a re-classification of \gv as WC8+O8\,III.

The radio mass loss rate of WR11 which was recently re-derived by Leitherer
et al.\ \markcite{lc97} (1997) has to be reduced by a factor of 2.3, 
which yields $\log \mdot=-4.55 \pm 0.16 \, \masl$.
This value is more typical of \mdot\ of WC9 stars which seem to show 
lower mass loss rates than WR stars of other types.
It may, however, be underestimated if the ionization in the outer
parts of the winds is lower than assumed (Leitherer et \markcite{lc97} al.).
Interestingly \mdot\ is now in very good agreement with results from
fits of ASCA X-ray observations with hydrodynamic wind-collision models 
(Stevens et al.\ \markcite{st96} 1996) allowing a distance independent
determination of \mdot $\sim$ \massloss{3.}{-5}.
Agreement between both methods weakens the requirement of important
structure (``clumping'') in the wind (cf.\ Stevens et al.\ \markcite{st96}
1996), which, if present, leads to an overestimate of the radio mass 
loss rate.

Depending on the adopted WR mass, the mass dependent mass loss rate 
formula of Langer \markcite{l89} (1989) for WC stars yields values 
between $\log \mdot=-4.78$ and $-3.80$ \masl. 
With the recently proposed relation of Schmutz \markcite{s97} (1997) 
the values are $\log \mdot = -5.32$ and $-4.54$ \masl. A better 
constraint on $M$ is clearly necessary for an accurate test of such 
a relation.

A 1.8 MeV $\gamma$-ray line emission of 2.3 to 
$5 \, 10^5$ cm$^{-2}$ s$^{-1}$ has been detected in the direction
of the Vela SNR (Diehl et al.\ \markcite{d95} 1995). 
From the $^{26}$Al yields of Meynet et al.\ \markcite{m97} (1997) the 
contribution of \gv to the $\gamma$-ray flux is estimated between 
1.5 and $2.6 \, 10^{-5}$ with our new distance measurement.
This values depends, however, quite strongly on the evolutionary scenario
of the WR star, which is still uncertain (cf.\ \S \ref{s_discussion}).

\section{Non-LTE model analysis for \gv stars}

From a line blanketed non-LTE analysis of the photospheric 
classification lines He\,{\sc i}\ $\lambda 4471$ and 
He\,{\sc ii}\ $\lambda 4541$ of the O star we derive
$\teff(O)=33500 \pm 1500$ K and $\log L(O)/\lsun=5.3 \pm 0.15$ for 
$M_V(O)=-5.13$.
Its Lyman continuum flux is $\log(Q_0)=48.7$ photon $s^{-1}$.
From these values we derive $M(O)=29 \pm 4$ \msun and an age of 
$\sim 4 \pm 0.5$ Myr from single star evolutionary models.
(Meynet et al.\ \markcite{mmssc} 1994).
With the age of the O star and the luminosity of the WC companion
(cf.\ below) single star evolutionary models yield an estimate of 
$M_{\rm ini} \sim 57 \pm 15 \msun$ for the initial mass of the WC8 
star.

{\footnotesize
\begin{table}[htb]
\caption{Stellar parameters for WR11 derived from non-LTE atmosphere models}
\centerline{
\begin{tabular}{lrrrrrrrrrr}
\\ \hline \\
ID & \tstar & \rstar  & $\log \mdot$ & $\log L$ & BC    & $\log Q_0$  \\
   & [kK]   & [\rsun] & [\masl]      & [\lsun]  & [mag] & s$^{-1}$ \\
\\ \hline
1    & 51. & 3.3 & -4.2 & 4.8 & -3.5 & 48.5 \\
2    & 61. & 1.9 & -4.3 & 4.7 & -3.4 & 48.4 \\ 
\hline
\end{tabular}
}
\label{ta_pars}
\end{table}
} 

{\footnotesize
\begin{table}[htb]
\caption{Summary of parameters for $\gamma^2$ Vel. Column 3 gives alternate
values for certain parameters}
\centerline{
\begin{tabular}{lrrl}
\\ \hline \\
Quantity  & value & alt.\ value & Ref. \\
\\ \hline
$\pi$ [mas] 	& 3.88 $\pm$ 0.53 & & SSG \\
$d=1/\pi$ [pc]	& $258^{+41}_{-31}$	& & SSG \\ 
$a$ [mas]	& $4.3 \pm 0.5$		& & HB70 \\
$P$ [days]	& 78.5002		& & NS80 \\
		& $\pm 0.0001$			 \\
$M_{\rm tot}$	& $29.5 \pm 15.9$	& & SSG  \\
$K({\rm O})$ [\kms]	& $40.9 \pm 6.5$	& $70 \pm 2$	& P83, M86 \\
$q$		& $0.36 \pm 0.09$	& $0.54 \pm 0.03$ & P83, M86 \\ 
$i$ [$^\circ$]	& $70 \pm 10$		& & ST87 \\
$e$		& $0.38	\pm 0.03$ & 0.40		& P83, M86 \\	
$M({\rm WR})$ [\msun] & $7.7 \pm 2.5$   & $19^{+7}_{-2}$  & P83, SM89 \\ 
$M({\rm O})$ [\msun]	& $22 \pm 5$ & $35^{+13}_{-3}$      & P83, SM89 \\
$\log L({\rm WR})_{\rm M-L}$ [\lsun]     & 5.06 & 5.69 & SSG \\	
\vinf({\rm WR}) [\kms]	& 1450&		& E94 \\
$\log \mdot({\rm WR})$ [\masl] & $-4.55 \pm 0.16$ & & SSG \\
$\log L({\rm WR})_{\rm NLTE}$ [\lsun] & $\sim$ 4.7 -- 4.8 & & SSG \\	 
$\teff({\rm O})$ [K]	& $34000 \pm 1500$ & & SSG \\
$\log L({\rm O})_{\rm NLTE}$ [\lsun]   & 5.3 $\pm 0.15$    & & SSG \\	 
$M({\rm O})_{\rm evol}$ [\msun]   & 29 $\pm 4$    & & SSG \\
age(O) [Myr]	& $4.0^{+0.8}_{-0.5}$ & & SSG \\	 
\hline
\\
\multicolumn{4}{l}{HB70: Hanbury Brown et al.\ \markcite{hb70} (1970)} \\
\multicolumn{4}{l}{NS80: Niemela \& Sahade \markcite{ns80} (1980)} \\
\multicolumn{4}{l}{M86: Moffat et al.\ \markcite{m86} (1986)} \\
\multicolumn{4}{l}{P83: Pike et al.\ \markcite{p83} (1983)} \\
\multicolumn{4}{l}{ST87: St.-Louis et al.\ \markcite{stn87} (1987)} \\
\multicolumn{4}{l}{SM89: Smith \& Maeder \markcite{sm89} (1989)} \\
\multicolumn{4}{l}{E94: Eenens et al.\ \markcite{e94} (1994)} \\
\multicolumn{4}{l}{SSG: this paper}
%
%
\end{tabular}
}
\label{ta_summary}
\end{table}
} 

A coarse analysis of the WC stellar parameters can be obtained
from a model grid following the general procedure
of Schmutz et al.\ \markcite{shw89} (1989). The non-LTE models used for 
this purpose include He and C with an abundance of C/He=0.25 by number.
The adopted terminal velocity is \vinf=1450 \kms\ (Eenens \& Williams
\markcite{e94} 1994). We use the observed equivalent widths of
\hei\ $\lambda$5876 and \heii\ $\lambda$5412 
(EW(5876)=$9.5 \pm 2.5$ \AA\ and EW(5412)=$4 \pm 1$ \AA) 
and correct these values for the contamination by the
light of the companion. 
The analysis then yields the temperature \tstar, radius \rstar, and \mdot.
From the uncertainties of the EWs and the fit procedure we estimate
an internal error of $\sim \pm 0.2$ dex on the luminosity $L$.
In Table~\ref{ta_pars} we list the parameters derived for the different 
assumptions for the magnitude difference between the components 
(cf.\ Table \ref{ta_mv}). 
We find that the \gv system provides approximately a factor of six 
less ionizing photons than \zp.

We have used a simpler model for carbon than the more elaborate
calculations by Koesterke \& Hamann \markcite{kh95} (1995). 
Test calculations with the parameters for the WC8 star WR135 derived by 
Koesterke \& Hamann \markcite{kh95} (1995) showed that our computations 
agree very accurately for the helium lines but we found less good 
agreement for the carbon lines.
Therefore, we have used only the helium lines for the spectroscopic 
diagnostic. We have also performed a coarse analysis using the grid of
Koesterke \& Hamann \markcite{kh95} (1995) for $\beta_C=0.2$ (C/He= 0.083 
by number). 
Although we get systematically lower temperatures we find that
$L$ is essentially identical.
We therefore conclude that the luminosity obtained from our model is 
consistent with the WC models from \markcite{kh95} Koesterke \& Hamann.
The parameters \tstar\ and \rstar\ depend quite strongly on 
assumptions about the velocity law and are therefore less reliable
(e.g.\ Schmutz \markcite{s96} 1996, \markcite{s97} 1997).

For the luminosity derived from the atmosphere models the M-L relation
gives $M(WR) \sim 5 \pm 1.5$ \msun, which is compatible with
the low mass determination for WR11.
On the other hand, if the mass is of the order of 19 \msun (as mostly
adopted in the literature) we are faced with a serious discrepancy
(up to an order of magnitude !) between $L$ derived from atmosphere and
evolutionary models.
The work of Howarth \& Schmutz \markcite{hs92} (1992) and Schmutz 
\markcite{s96} \markcite{s97} (1996, 1997) previously revealed differences 
for some WN stars. According to 
Schmutz \markcite{s97} \markcite{s97} (1996, 1997) this is most likely 
due to systematic errors in the non-LTE model atmosphere, and it may lead 
to underestimates of $L$ by typically 50 to 300 \%.
We conclude that for the low mass determination of WR11 the 
theoretical M-L relation for hydrogen-free WR stars agrees 
within the errors with $L$ derived from atmosphere models.

\section{Discussion}
\label{s_discussion}
The distance and the origin of the Gum nebula, the spatial distribution 
of important  objects in this region (Gum nebula, Vela R2 and Vela OB
associations, \zp, \gv, the Vela SNR etc.) and their possible relation 
are still relatively poorly known and controversial 
(cf.\ e.g.\ Brandt et al.\ \markcite{b71} 1971, Bruhweiler et al.\ 
\markcite{b83} 1983, Franco \markcite{f90} 1990, Sahu \markcite{s92} 1992, 
Sahu \& Blaauw \markcite{sb93} 1993, Oberlack et al.\ \markcite{o94} 1994,
Fitzpatrick \& Spitzer \markcite{fs94} 1994, Aschenbach et al.\ 
\markcite{a95} 1995).
Combining recent data (including additional HIPPARCOS parallaxes and
proper motions) with the most elaborate study of this region
done by Sahu \markcite{s92} (1992) may lead to a refined picture. 
While such a task is clearly beyond the scope of the present paper 
we shall in the following briefly discuss a few questions related to 
\zp and \gv.

Based on its proper motion Sahu \markcite{s92} (1992) suggested that the 
runaway star \zp originates from the more distant young Vela R2 association.
Although the HIPPARCOS distance to \zp is lower than the $\sim$ 730 pc
suggested by Sahu, her scenario remains valid provided the distance
to Vela R2 is less than the adopted 800 pc\footnote{Adopting our one
sigma error on $d$ of \zp the distance of Vela R2 must be $\la$ 580 
pc.}, which is compatible with the distance estimate of
Liseau et al.\ \markcite{l92} (1992) and the HIPPARCOS parallax of two 
presumed members (HD 76534: 2.43 $\pm$ 1.30, HD 76838: 3.39 $\pm$ 0.90 mas). 
Evolutionary scenarios for \zp have been presented by van Rensbergen 
et al.\ \markcite{vr96} (1996). 
It would be highly desirable to include the effects of the rapid rotation
of \zp in future calculations.

We obtained additional HIPPARCOS parallaxes for 5 presumed Vela OB2 
members (HD 63922: 1.66 $\pm$ 0.53, HD 64740: 4.53 $\pm$ 0.52, HD 68657:
4.48 $\pm$ 0.49, HD 70930: 2.16 $\pm$ 0.57, HD 72485: 3.10 $\pm$ 0.52,
all values in mas) for which Sahu \markcite{s92} (1992) derived proper 
motions and statistical parallaxes. 
With the exception of HD 68657 (and \gv) her values are within 
1 $\sigma$ of the HIPPARCOS measurements. 
The unexpectedly large annual parallax of \gv lies thus well within 
the range covered by other stars which, as shown by Sahu \markcite{s92} 
(1992), are likely members of the association Vela OB2.
On the basis of our limited data (no new proper motions) there is
no evidence against the membership of \gv to this group of objects.
Given the very small number of objects and the apparently large distance 
spread determined from the annual parallaxes, this association seems, 
however, to be quite loosely defined.
Despite these uncertainties an age of $\sim$ 20 to 30 Myr (Eggen 
\markcite{e80} 1980, Sahu \markcite{s92} 1992) is usually estimated. 
For the objects mentioned above we obtain individual ages between 
$\sim$ 6-30 Myr from Geneva photometry and the HIPPARCOS data. 
For the majority we favor $\sim$ 10 Myr, which is also supported by 
recent kinematic observations of the Vela Shell (Churchwell et al.\ 
\markcite{c96} 1996).

Single star evolution models fail to explain the formation of the 
\gv system for ages $\gtrsim$ 6 Myr (cf.\ Schild \& Maeder \markcite{sm84} 
1984). A young age is, however, compatible with our estimate for the O star.
An older age of $\sim$ 10 Myr as typical for other Vela OB2 stars
would exclude a large total mass of the system ($\sim 50-60 \, \msun$).
On the other hand $M_{\rm tot} \sim 30 \, \msun$ 
might be compatible with 10 Myr if WR11 evolved from a
intermediate-mass progenitor through mass transfer in the binary system.
However, the relatively large eccentricity of \gv may 
speak against such a scenario.
Regrettably, the present data does not allow to draw more definite 
conclusions about the formation of the \gv system.

\acknowledgements{We thank Drs. A. Kaufer, J. Schweickhardt,
O. Stahl, and B. Wolf for 
the spectra of \gv obtained with HEROS at the ESO 0.5m telescope,
and Dr. V. Niemela for discussions on spectral classification.
DS wishes to thank Georges Meynet, Jean-Claude Mermilliod, Claus Leitherer 
and J\"urgen Kn\"odlseder for useful discussions.
Travel support from the Geneva Observatory and an invitation from
the Universit\'e Paul-Sabatier in Toulouse, where part of this work
was carried out, are kindly acknowledged. 
DS receives a fellowship from the Swiss National 
Foundation of Scientific Research and partial support from the 
Directors Discretionary Research Fund of the STScI. }



\end{document}